\documentstyle[preprint,aps]{revtex}
\begin{document}
\bibliographystyle{unsrt}
\def\Journal#1#2#3#4{{#1} {\bf #2}, #3 (#4)} 
\def\NCA{\em Nuovo Cimento} 
\def\NIM{\em Nucl. Instrum. Methods} 
\def\NIMA{{\em Nucl.Instrum. Methods} A} 
\def\NPB{{\em Nucl. Phys.} B}
\def\PLB{{\em Phys. Lett.}B} 
\def\PRL{\em Phys. Rev. Lett.}
\def\PRD{{\em Phys. Rev.} D}
\def\ZPC{{\em Z. Phys.} C}
\def\MPLA{{\em Mod. Phys.Lett} A}
\def\ANP{\em Ann. Phys.} 
\def\PTP{\em Prog. Theor. Phys.} 
\def\PR{\em Phys. Rep.} 
\def\st{\scriptstyle}
\def\sst{\scriptscriptstyle}
\def\mco{\multicolumn}
\def\epp{\epsilon^{\prime}}
\def\vep{\varepsilon}
\def\ra{\rightarrow}
\def\ppg{\pi^+\pi^-\gamma}
\def\vp{{\bf p}}\def\ko{K^0}
\def\kb{\bar{K^0}}
\def\al{\alpha}
\def\ab{\bar{\alpha}}
\def\be{\begin{equation}}
\def\ee{\end{equation}}
\def\bea{\begin{eqnarray}}
\def\eea{\end{eqnarray}}
\def\CPbar{\hbox{{\rmCP}
\hskip-1.80em{/}}}

\preprint{\begin{tabular}{c}
\hbox to\textwidth{February 1998 \hfill BROWN-HET-1109}\\
[-10pt]
\hbox to\textwidth{ \hfill BROWN-TA-554}\\
[-10pt]
\hbox to\textwidth{ \hfill KIAS-P98003}\\[-10pt]
\hbox to\textwidth{ \hfill hep-ph/9802xxx}\\ [-10pt]
\end{tabular}}
\draft
\title{New Class of Quark Mass Matrices and the Flavor Mixing
Matrix\footnote{Supported in part by the 
USDOE contract DE-FG02-91ER 40688-Task A and presented by one of us (SKK) at the APCTP Workshop: {\it Pacific Particle Physics Phenomenology}, Seoul, Korea, 31 October - 2 November, 1997.}} 
\author{Kyungsik Kang}
\address{\it
Department of Physics, Brown University,
Providence, RI 02912, USA}
\author{Sin Kyu Kang} 
\address{\it
School of Physics, Korea Institute for Advanced Study, Seoul, Korea}
\maketitle
%
%

\begin{abstract}
We discuss a new general class of mass matrix ansatz that respects
the fermion mass hierarchy and {\it calculability} of the flavor mixing matrix.
This is a generalization  of the various specific forms 
of the mass matrix that is obtained
by successive breaking of the maximal permutation symmetry.
By confronting the experimental data, a large class of the mass 
matrices are shown to survive, while certain specific cases are 
phenomenologically ruled out.
\end{abstract} 
\begin{center}

\end{center}

%

The flavor mixing and fermion masses and their hierarchical 
patterns remain to be one of the basic problems in particle physics.
Within the standard model(SM), all masses and flavor mixing angle  are free 
parameters and no relations among them are provided.
%
As an attempt to derive relationship between the quark masses and flavor mixing
hierarchies, mass matrix ansatz was suggested about two decades ago 
\cite{wein}.
This in fact reflects the {\it calculability} \cite{wein,cal}
of the flavor mixing angles in terms of the quark masses.
Of several ansatz proposed, the canonical mass matrices of the 
Fritzsch type \cite{wein,test} have been generally assumed to predict the 
entire Kobayashi-Maskawa (KM) matrix \cite{km} or the Wolfenstein mixing
matrix \cite{wolf}.
Though the Fritzsch texture \cite{wein,cal} is attractive 
because of its maximal {\it calculability},
it predicts a top quark mass to be no larger than 100 GeV and thus is 
ruled out \cite{test}.
Thus, the next move is to modify the Fritzsch mass matrix by introducing
just one more parameter but by maintaining the {\it calculability} property 
\cite{kk0}.
In what follows, we will discuss a possible modification of the
Fritzsch mass matrix.

A natural choice for the next nonvanishing entry in the Fritzsch matrix is 
the (2,2) element.
Although such type of mass matrices has been studied in the literature
\cite{kaus,fr2,wu,xing},
we will show that they
can be identified as special cases of the general form in which
(2,2) and (2,3) elements are related in a particular way.
The general form can be achieved by breaking the 
democratic flavor
symmetry $S(3)_L \times S(3)_R $ successively down to
$S(2)_L \times S(2)_R$ and then to $S(1)_L \times S(1)_R$.

As is well known, the $3\times 3$ ``democratic mass matrix ",
\begin{eqnarray}
   \frac{c}{3}~\left( \begin{array}{ccc}
               1 & 1 & 1 \\
               1 & 1 & 1 \\
               1 & 1 & 1  \end{array} \right),
\end{eqnarray}
exhibits the maximal $ S(3)_{L} \times S(3)_{R} $ permutation symmetry.
This can be achieved by breaking  the chiral symmetry $U(3)_L \times
U(3)_R$ to $S(3)_L \times S(3)_R$, where $U(3)$ is the symmetry group 
connecting the three generations \cite{kaus,fr3}.
One may say that the scale of this chiral symmetry breaking is the electroweak
symmetry breaking scale at which the third generation quarks get masses.
Indeed, one can see this by making
unitary transformation of (1) with the help of 
$ U=(u_1^{T}, u_2^{T}, u_3^{T})$, where $u_1=(\frac{1}{\sqrt{2}},
\frac{1}{\sqrt{6}}, \frac{1}{\sqrt{3}}), 
u_2=(-\frac{1}{\sqrt{2}},  \frac{1}{\sqrt{6}}, 
\frac{1}{\sqrt{3}})$ and $u_3=(0, -\frac{2}{\sqrt{6}}, \frac{1}{\sqrt{3}} )$.
In order to account for the hierarchical pattern of the second and first
generation quark masses,  one has to break the $S(3)_{L} \times S(3)_{R} $ 
symmetry successively in two stages to $S(2)_L \times S(2)_R$ and
$S(1)_L \times S(1)_R$.
This can be achieved by adding the following two matrices to (1):
\begin{eqnarray}
               \left( \begin{array}{ccc}
               0 & 0 & a \\
               0 & 0 & a \\
               a & a & b \end{array} \right),
 ~~ \qquad
         d ~\left( \begin{array}{ccc}
               1 & 0 & -1 \\
               0 & -1 & 1 \\
              -1 & 1 & 0  \end{array} \right),
\end{eqnarray}
where the parameters $(a,b)$ and $d $ are responsible for the breakdown
of $S(3)_{L}\times S(3)_{R}$ and $S(2)_{L}\times S(2)_{R}$ symmetries,
respectively.
It is also reasonable to anticipate that this two-stage breaking happens 
at around 1 GeV, the chiral symmetry breaking
scale, in view of the proximity of the second and first generation quark
masses compared to the third generation quarks.
Since the evolution from the electroweak scale to 1 GeV scale can not
alter the `` democratic " pattern of the mass matrix,
the resulting mass matrix can be regarded as the one at 1 GeV scale.
%
Then the mass matrix in the hierarchical basis reduces
after the unitary transformation with $U$ to, 
\begin{eqnarray}
   M_H = \left( \begin{array}{ccc}
               0 & A & 0 \\
               A & D & B \\
               0 & B & C  \end{array} \right),
& ~~ \qquad
\end{eqnarray}
where $A=\sqrt{3}d, D=-\frac{2}{3}(2a-b), B=-\frac{\sqrt{2}}{3}(a+b)$ and
$C= \frac{1}{3}(4a+b)+c $.  

Note that in order to get a hermitian mass matrix instead of (3), one can use
the following two matrices in the place of (2),
\begin{eqnarray}
          \left( \begin{array}{ccc}
            p & p  & a+q \\
            p & p  & a+q \\
            a+q^{\ast} & a+q^{\ast} & b-2p 
               \end{array} \right), 
 ~~ \qquad
         d~\left( \begin{array}{ccc}
               \cos \sigma & -i\sin \sigma & -e^{-i\sigma} \\
               i\sin \sigma & -\cos \sigma &  e^{-i\sigma} \\
               -e^{i\sigma} & e^{i\sigma} & 0 
                \end{array} \right),
\end{eqnarray}
where $p=\frac{4}{9}(a+b)\sin^2 \frac{\delta}{2} $ and $q=p(1+i\frac{3}{2}
\frac{e^{i\delta/2}}{\sin \frac{\delta}{2}}) $.
Then, after the unitary transformation with $U$,
the (1,2) and (2,3) elements in $M_H$ become $A e^{-i\sigma}$ and
$B e^{-i\delta}$ respectively.
However, since  only one phase factor is sufficient to describe the 
CP-violation in the SM containing three family generations of 
quarks, we may 
introduce only one phase factor in the  
hermitian matrix $M_H$ i.e., only to the (1,2) and (2,1) elements.
In this way, a hermitian mass matrix of the type (3), with complex
elements at (1,2) and (2,1), 
can be obtained. 

At a glance, the matrix $M_H$ contains four independent parameters even in
the case of real parameters so that the {\it calculability} is lost.
However, one can make additional ansatz to relate $a$ to $b$, 
so that $a=kb$ in general, with the same ratio parameter $k$ for both the 
up- and down-quark sectors, so as to maintain the {\it calculability}.
Then, the (2,2) element is related to (2,3) element by
$w\equiv B/D = (k+1)/\sqrt{2}(2k-1) $ in the
hierarchical mass eigenstates. Moreover, various specific mass matrices 
proposed 
by others can be identified as a special case of the new mass matrix
i.e., $w=\frac{5}{3}~(k=0.9)$ for Ref. 7
, $w=-\frac{1}{\sqrt{2}}~(k=0)$ for Fritzsch {\it et al}. \cite{fr2}, 
$w=\pm 2\sqrt{2}~(k=\frac{5}{7} ~\mbox{or}~ \frac{1}{3}) $ for 
Ref. 10
and $w=\sqrt{2}~(k=1)$ for Ref. 11.
The case of $k=\frac{1}{2} $ reduces to the old Fritzsch type with
$D=0 $.

The next step is then  to constrain $k$ for the general class of  mass matrix 
by confronting the experiments for consistency. 
The mass matrix $M_H$ of the type (3) can be brought to a diagonal form by 
a biunitary transformation,
$U^{(u,d)}_L M^{(u,d)}_{H} U^{(u,d)^{\dagger}}_R =
 diag[m_{u,d}, m_{c,s}, m_{t,b}]$.
Since $U_L M_H U_L^{\dagger}$ and
$U_R M_H U_R^{\dagger}$ are diagonal, $U_LU_R^{\dagger}\equiv K$ is 
again diagonal.
It turns out in general that, because of the empirical mass
hierarchy $m_1 \ll m_2 \ll m_3 $,
$K=diag[1,-1,1]$ irrespective
of the sign of $D$ and $K=diag[-1,1,1]$ only for positive $D$. 
This point was not clearly understood in previous works 
\cite{fr2,wu,xing,gupta}.
%
The parameters $A,B,C$ and $D$ can be expressed in terms of the quark
masses.
In view of the hierarchical pattern of the quark masses, it is natural
to expect that $A \ll |D| \ll C $, and then the case of $K=diag[1,-1,1]$
for positive $D$ can be excluded if the same ratio parameter $w$ is required
for both up- and down-quark sectors.
Otherwise, the masses of the second family could be unacceptably large.

{\it The Case $K=diag[-1, 1, 1]$}:  
The hermitian matrix  can be written as $M^{(u,d)}_{H} = P^{(u,d)} M_r^{(u,d)} {\tilde P^{(u.d)}}$, where
$P^{(u,d)}=diag[\exp(-i\sigma^{(u,d)}), 1, 1]$,
and the real matrix $M_r^{(u,d)}$ can be diagonalized by a real
orthogonal matrix $R^{(u,d)}$ so that
$R^{(u,d)}M_r^{(u,d)}{\tilde R^{(u,d)}} = 
diag[-m_{(u,d)}, m_{(c,s)}, m_{(t,b)}].$
Then 
the flavor mixing matrix is given by $V=U^{(u)}_L U^{(d)^{\dagger}}_{L}=
{\tilde R^{(u)}} P R^{(d)}$
where $P=diag [e^{i\sigma }, 1, 1]$ with 
$\sigma = \sigma^{(u)} - \sigma^{(d)}$.

>From the characteristic equation for the $M_r$, the mass matrix $M_r$
can be written by
\begin{eqnarray}
   M_r = \left( \begin{array}{ccc}
               0 & \sqrt{\frac{m_1 m_2}{1-\frac{\epsilon}{m_3}}} & 0 \\
               \sqrt{\frac{m_1 m_2}{1-\frac{\epsilon}{m_3}}} & 
               m_2-m_1+\epsilon & w(m_2-m_1+\epsilon) \\
               0 & w(m_2-m_1+\epsilon ) &
               m_3-\epsilon  \end{array} \right)
\end{eqnarray}
in which the small parameter $\epsilon$ is related to $w$, i.e.,
$w \simeq \pm \frac{\sqrt{\epsilon m_3}}{m_2}
\left(1+\frac{m_1}{m_2}-\frac{m_2}{2m_3}\right)$,
whose range is to be determined from the experiments.
Then, we can obtain analytic expressions for the flavor mixing matrix
$V$ which gives in the leading approximation 
\begin{eqnarray}
|V_{us}| &\simeq &\left| \sqrt{m_d/m_s}\exp{(i\sigma)}-
                        \sqrt{m_u/m_c}\right|, \\
|V_{cb}| &\simeq & \left| w
           \left(m_s/m_b-m_c/m_t\right)\right|, \\
|V_{ub}|/|V_{cb}| &\simeq & \sqrt{m_u/m_c}, ~~~
|V_{td}|/|V_{ts}| \simeq \sqrt{m_d/m_s}.
\end{eqnarray}
Since the second term of $|V_{cb}|$ is negligible compared to the first term,
it is easy to examine the range of $w$ for which
$|V_{cb}|$ is compatible with experiments.
Using the quark masses given in Ref. 14
and the experimental value $|V_{cb}| = 0.036 - 0.046 $ \cite{data},
(7) leads to $1.01\leq |w| \leq 2.02$ so that
$0.82\leq k \leq 1.31$ if $w>0$ and
$0.11\leq k \leq 0.28$ if $w<0$
in the leading approximation, which is close to the exact result
 $0.97\leq |w| \leq 1.87$ so that
$0.85\leq k \leq 1.36$ if $w>0$ and
$0.10\leq k \leq 0.26$ if $w<0$.
Note that $\epsilon \simeq O(m_1)$ for the allowed range of $k$ and $w$.

Next, we examine if this range of $w$ preserves the consistency
with experiments for other KM elements.
Since several KM elements depend on the phase factor $\sigma$,
we have to determine the allowed range of the phase factor first.  
We see from  (6) that $|V_{us}|$ depends 
on the phase factor $\sigma $, while independent of $w$.
Using the experimental value $|V_{us}| \simeq 0.219 - 0.224$ \cite{data} 
the allowed range of $\sigma $ turns out to be 
$26^{\circ}-111^{\circ}$. 
The exact numerical result gives
$39^{\circ} \leq \sigma \leq 117^{\circ}$.
In addition we find that all other KM elements are in good agreement with
experiments for the above ranges of $w$ and $\sigma $.

{\it The Case $K=diag[1, -1, 1]$}:  For a negative $D$, the real symmetric 
matrix $M_r^{(u,d)}$ can be diagonalized as 
$R^{(u,d)}M_r^{(u,d)}{\tilde R^{(u,d)}} 
= diag[m_{(u,d)}, -m_{(c,s)}, m_{(t,b)}]$, 
thus reversing the signs of $m_1$ and $m_2$ in (5). 
Following the similar analysis as in the previous case, we get 
$1.14\leq |w| \leq 2.76$ so that
$0.72\leq k \leq 1.17$ if $w>0$ and
$0.14\leq k \leq 0.33$ if $w<0$,
and the same range of $\sigma$ as in the previous case
in the exact numerical calculation,
while we find the same result of $w$ and $\sigma$
as in the previous case in the leading approximation.
Consequently
the ansatz adopted by Fritzsch {\it et al}. \cite{fr2},
corresponding to $k=0$,
is not consistent with
experimental data of $V_{cb}$ and the ansatz adopted by
Ref. 10
, corresponding to $w^2=8$, is 
slightly beyond the upper bound of the allowed $w$.
Finally, we note that the predicted ratio $|V_{ub}|/|V_{cb}|$ ($\leq 0.07$)
tends to be on the low side of (but consistent with ) the present experimental 
range, 
$|V_{ub}|/|V_{cb}|=0.08\pm 0.02$ \cite{data} or $0.08 \pm 0.016$ \cite{ratio1}.

\end{document}